\documentclass[conference]{IEEEtran}
\IEEEoverridecommandlockouts
\usepackage{cite}
\usepackage{amsmath,amssymb,amsfonts}
\usepackage{algorithmic}
\usepackage{graphicx}
\usepackage{textcomp}
\usepackage{xcolor}
\usepackage{paralist}
\def\BibTeX{{\rm B\kern-.05em{\sc i\kern-.025em b}\kern-.08em
    T\kern-.1667em\lower.7ex\hbox{E}\kern-.125emX}}
\begin{document}

\title{Towards predictive crowd based transport infrastructure monitoring system}

\author{\IEEEauthorblockN{1\textsuperscript{st} Fatjon Seraj}
\IEEEauthorblockA{\textit{Pervasive Systems Group} \\
\textit{University of Twente}\\
Enschede, the Netherlands \\
f.seraj@utwente.nl}
}

\maketitle

\begin{abstract}
To be able to measure relevant data for transport infrastructure monitoring and to obtain maintenance indicators in a crowd sensing-based fashion, a set of requirements (both from hardware and software point of views) needs to be satisfied. Heterogeneity of smartphones and various uncertainties associated with the mainstream \textit{off-the-shelf} hardware combined with the fact that inexperienced participants will take an active part in the data collection process necessitates that the system accounts for dynamicity, uncertainty, unreliability, and heterogeneity of the environment, participants, and technology.
Modern Software Development Kits (SDK) and Application Programming Interfaces (API) provided by the mobile Operating Systems allow us to build applications for the smartphones ecosystem and to interact with components, services, and applications. They also provide specifications for hardware components and modules (e.g. sensors, location, and networking modules).
Being a system implemented on a heterogeneous and dynamic smartphone platform, our crowd sensing-based transport infrastructure monitoring system satisfies a number of hardware and software requirements. Hence, we propose a detailed and rigorous methodology to research the requirements and describe the overall system architecture including its functionalities and services.
\end{abstract}

\section{Introduction}
\label{ch_intro}
Transportation infrastructure is in the core of our civilization. It allows us to travel faster and further, to explore vast territories, and to meet each other. We are continuously on the move and the need to be mobile all the time has lead to important innovations that have shaped our current transport infrastructure. 
Everything started with narrow trails our ancestors used for hunting. As animals got domesticated, the trails got wider to accommodate the  traffic. The invention of the wheel allowed people to increase their traveled distances and transported load. However, to be able to ride smoothly, the wheeled carts required flatter roads. 

Many studies and surveys are made on the topic of roadway deficiencies and their impact on safety and economy~\cite{Miller:2009te}. 
With the advancement of the automotive industry, transport infrastructure improved to allow higher travelling speeds. 
However, higher speeds cannot tolerate road imperfections and defects. To ensure a high safety standard for passengers, vehicles, and goods the maintenance should be adequate and up-to-date. 

Transport infrastructure may wear and its condition deteriorate over time due to various factors related to, among other things, their location, load/traffic, weather, engineering solutions, and materials.
Road anomalies such as potholes can affect the driving experience and the overall state of the vehicle.

Developed countries monitor their infrastructure through specialized systems called Pavement Management Systems (PMS). 
The Pavement Management System consists of a set of tools that assists authorities to develop cost effective strategies for  evaluating and maintaining road pavements in working condition~\cite{kinney_development_1986}. PMSs have two major components, i.e., (i) a complete database, which stores all available information about the current and historical pavement condition, road transport infrastructure, and traffic, and (ii) a set of statistical and predictive tools to allow authorities to evaluate existing and future pavement conditions and to identify and to prioritize infrastructural investments. 

In the Netherlands since the introduction of PMSs in early 1989, {75\%} of all of the local authorities, municipalities, and provinces, have utilized a PMS~\cite{schut_responsible_2000}. Developing countries often lack this kind of technology and the \textit{know-how}.

To be able to measure relevant data for transport infrastructure monitoring and to obtain maintenance indicators in a crowd sensing-based fashion, a set of requirements (both from hardware and software point of views) needs to be satisfied. Heterogeneity of smartphones and various uncertainties associated with the mainstream \textit{off-the-shelf} hardware combined with the fact that inexperienced participants will take an active part in the data collection process necessitates that the system accounts for dynamicity, uncertainty, unreliability, and heterogeneity of the environment, participants, and technology.
Smartphone echo-system is dominated by the major Operating System providers, namely, iOS and Android from Apple and Google, which provide Software Development Kits (SDK) and Application Programming Interfaces (API) that allow developers to build applications for the smartphones ecosystem and to interact with components, services, and applications. They also provide specifications for hardware components and modules (e.g. sensors, location, and networking modules). Whereas iOS maintains uniformity of sensor components overall its smartphones, the quality of Android devices varies by a myriad of manufacturers covering the whole price spectrum.
Being a system implemented on a heterogeneous and dynamic smartphone platform, our crowd sensing-based transport infrastructure monitoring system needs to satisfy a number of hardware and software requirements. Hence, a detailed and rigorous methodology to research the requirements and describe the overall system architecture including its functionalities and services.
\section{Sensing requirements}
Most of the signals encountered in science and engineering are analog in nature. That means that these signals are functions of a continuous variable, such as time or space, and usually take on values in a continuous range~\cite{ProakisManolakis200604}.
Physical signals are measured by domain specific devices called sensors. For example, sound and vibration signals are recorded by a microphone and an accelerometer, respectively. 
However, what is recorded by these devices is only an approximation of the true event due to the constrains and errors associated with the sensors themselves. 
Nonetheless, if the sensor errors are less than the required accuracy they can successfully observe the given event. Our system requires the following sensors, which majority of modern smartphones fulfill them:

\subsection{Inertial sensors}
Accelerometer is an inertial sensor capable of measuring the vector quantity of magnitude and direction of the proper acceleration. In other words, it measures the rate of change of velocity of the smartphone with respect to time. Currently accelerometer sensors of smartphones are primarily used for re-orienting the screen and its content based on the sensed orientation of the device.


Instantaneous acceleration is measured over an infinitesimal time interval and, as expressed below, is the limit of average acceleration when the time interval approaches zero. This equals the derivative of velocity with respect to time.
\begin{equation}
   a=\lim_{\Delta t \to 0} \frac{\Delta v}{\Delta t}=\frac{dv}{dt} 
\end{equation}

Velocity is the derivative of displacement $r$ with respect to time and is expressed as:
\begin{equation}
a=\frac{dv}{dt}=\frac{d}{dt}\frac{dr}{dt}=\frac{d^{2}r}{dt^{2}}   
\end{equation}

Linear acceleration denoted by $la$ is the measured acceleration $a$ without the gravity component $g$. 
Android provides the linear acceleration as the output of a \textit{synthetic sensor}, i.e., a virtual sensor, being calculated based on the input of acceleration, gyroscope, and magnetic sensor. In fact Android implements a Kalman filter to reduce the errors introduced by the sensor noise and provides output of a number of \textit{synthetic sensors}.
When smartphones are equipped only with the accelerometer sensor, the linear acceleration $la$ can be calculated by first calculating the gravity vector $g$.
Considering the fact that gravity is somehow constant (i.e., not changing very fast), it can be approximated applying a low-pass filter on $a$ with response $\alpha$:
\begin{equation}
    \begin{split}
     \alpha=\frac{t}{t+dt}  \enspace |\enspace
      g=\alpha * g + (1-\alpha)*a \enspace|\enspace
        la=a-g 
    \end{split}
\end{equation}

This implies that availability of accelerometer sensor alone is enough to have an estimation of the linear acceleration and angle of rotation. However, availability of the gyroscope and magnetometer sensors will increase the accuracy of displacement and rotation utilizing  sensor fusion techniques.

Gyroscope is an inertial sensor capable of measuring the change of rotational (angular) velocity $\omega$, the speed at which the smartpphone is rotating. In other words, it measures the angular$\phi$ displacement with respect to time.
\begin{equation}
\begin{split}
\omega =\frac{d\phi}{dt} \quad \quad
\vec{\phi}=\int \vec{\omega} dt    
\end{split}
\end{equation}

Gyroscope is used in the smartphones to increase the accuracy of the device rotations, helping developers to build applications that respond more accurately to the user input.  


\subsection{GPS sensor}
Speed of travel affects the measurements in both frequency and amplitude, therefore becomes necessary to know the speed in order to reduce its dependency. Localization of the measured infrastructure segments is another crucial requirement. 

Using a device equipped with GPS (\textit{\textbf{G}lobal\textbf{P}ositioning \textbf{S}ystem}), one may assume that speed and position will always be available. However, this is not the case with the current GPS chips found in today smartphones. 
GPS sensors suffer from the so-called TTFF (\textit{\textbf{T}ime \textbf{t}o \textbf{F}irst \textbf{F}ix}) delay. In order to ensure short TTFF, availability of a clear line of sight and a stable position are essential. This is not always the case, especially for a travelling vehicle. In smartphones, the TTFF is improved by introducing a technology called A-GPS (Assisted GPS)~\cite{AGPS:_2002}. A-GPS requires an active data connection with a mobile network operator to receive a preloaded list of available GPS satellites for a given location location. In addition to shortening the TTFF, A-GPS also improves GPS location accuracy by providing the cell tower localization. 



\section{Services}
\begin{figure}[ht]
	\centering
	  \includegraphics[keepaspectratio=true, width =0.46\textwidth]{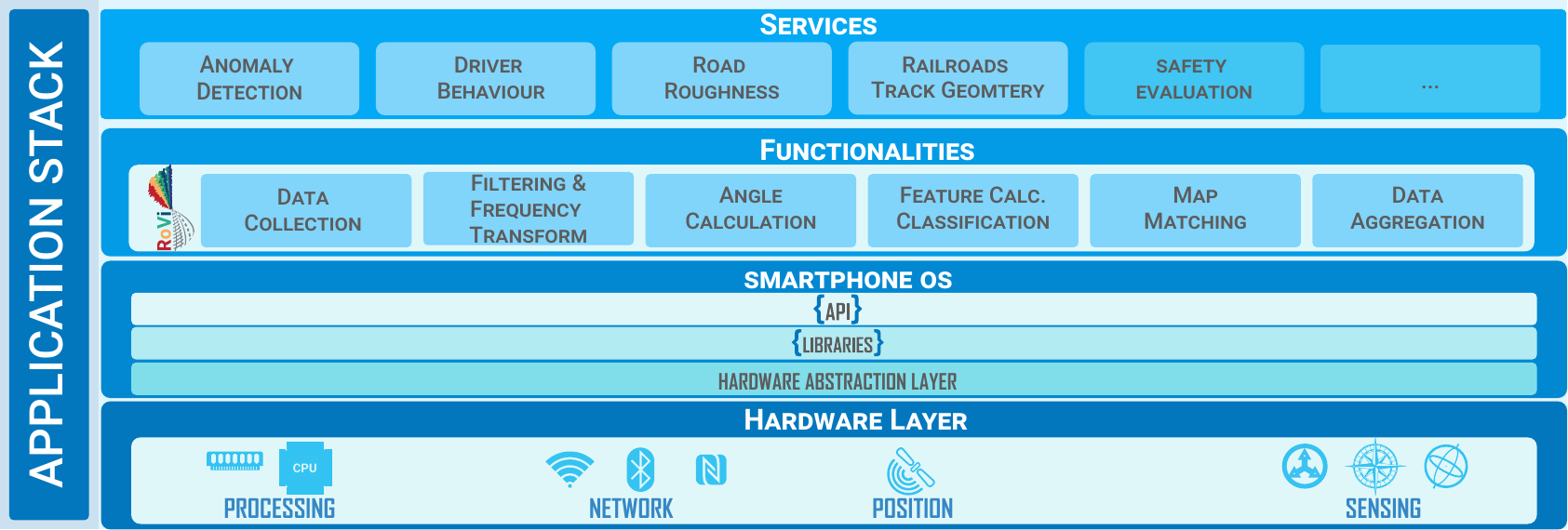}
				\small\caption{Smartphone based transport infrastructure monitoring application stack}
				\label{fig:api}
	\end{figure}

Monitoring different transport infrastructure types requires different services and maintenance indicators. 
This means that all these services need to be part of a multi-purpose framework that various services can be implemented individually for domain-specific applications. 
Figure~\ref{fig:api} demonstrates a synopsis of our proposed framework system and corresponding application stack representing various services.
The framework is proposed as a skeleton providing a set of methods and techniques to build the required services and applications related to the transport infrastructure monitoring. The following chapters will provide more details regarding these techniques and algorithms as well as their evaluation in different application scenarios. In what follows, we describe these services and their underlying principles briefly.

Our system provides three main services related to the road transport infrastructure monitoring, namely anomaly detection, driver behavior analysis, and road roughness analysis. Railroad track geometry service handles the calculation of track geometry indicators 
\subsubsection{Anomaly detection}
Anomaly detection refers to techniques that identify anomalous data observations or events which diverge from expected patterns, behaviors, or events in a dataset. 
These non-conforming patterns are often referred to as anomalies, outliers, discordant observations, exceptions, aberrations, surprises, peculiarities, or contaminants in different application domains~\cite{chandola_anomaly_2009}.
Anomalies can be classified into three categories~\cite{chandola_anomaly_2009}:
\begin{inparaenum}
    
    \item Point anomalies - When individual data instances differ from the rest of the data. Example of such anomalies are the potholes, speed-bumps, manholes etc.
    
    \item Contextual anomalies - When data instances are anomalous only in a specific context. For example consider a driver forced to swerve around a pothole. This swerve can be considered anomalous in comparison with other type of turns, because the driver is expected to drive straight in that particular segment of roads.
    
    \item Collective anomalies - When the data instances \textit{per se} may not be anomalous but their evolution over time or space makes them anomalous. An example is the continuous monitoring of the road pavement, although the pavement does not exhibit any visual deterioration, the pavement layer in time wears off and the deterioration trend over time shows the anomalous segments.
\end{inparaenum}

Anomaly detection in the context of transport infrastructure monitoring refers to identification of all unexpected phenomena, situations, and events faced on these segments/pathways.
From a naive-reasoning point of view, anomalies on the transport infrastructure generate excessive vibration when a vehicle runs over them. These anomalous events are expected to exhibit high amplitude vibrations and can be classified as \textit{point anomalies}.
This hypothesis is, however, only true if the entire  vehicle fleet rolls on the the infrastructure with the same speed. 
However, since this is not always the case, monitoring the amplitude of the vertical vibration alone is not enough to correctly detect and classify them.
Accurate anomaly detection requires \textit{a priori} knowledge (labels) about type of frequently occurring anomalies, their representation in the dataset, as well as transport infrastructure segments corresponding to both normal and anomalous situations.

\subsubsection{Driver behavior analysis}
Driver behavior on the road is characterized by drivers attitude towards the speed and the turns. 
Road engineers divide the road curves into two categories, i.e., horizontal and vertical curves. Horizontal curves change the direction of the vehicle from left to right or right to left, while vertical curves or slopes change the direction of the vehicle from up to down or down to up. A maneuver on the road is in essence a curve or a set of curves, therefore the focus of identifying maneuvers is on the horizontal changes of vehicle direction. 

A driver will follow a curve in the following cases: (i) when the vehicle follows a turn on the road, (ii) when the vehicle changes the road lanes, (iii) when the vehicle overtakes another vehicle, a pedestrian, or a cyclist, (iv) when the vehicle drives on a curvy road, (v) when the vehicle aims to avoids a (risky) anomaly on the road pavement.

Curves can be permanent in the form of road turns or temporal in the form of swerves. Temporal curves are caused by the driver decision to overcome temporal situations on the road, for instance overtaking events or swerving to avoid objects, debris on the road, dead animals, tree branches, etc.

Our driver behavior analysis service based on the algorithm described in \cite{Seraj:angle}, calculates all vehicles jaw axis angles and classifies the curves into different categories based on the angle of the vehicle swerve and the complexity and signature of the swerve elements.
The information inferred from the driving behavior analysis service can be used to enhance the detection of infrastructure anomalies, to infer hazardous infrastructure segments, to study the behavior of young drivers, among other things. 


\subsubsection{Road roughness analysis}

\begin{figure}[ht]
  \begin{center}
 \includegraphics[keepaspectratio=true, width =0.46\textwidth]{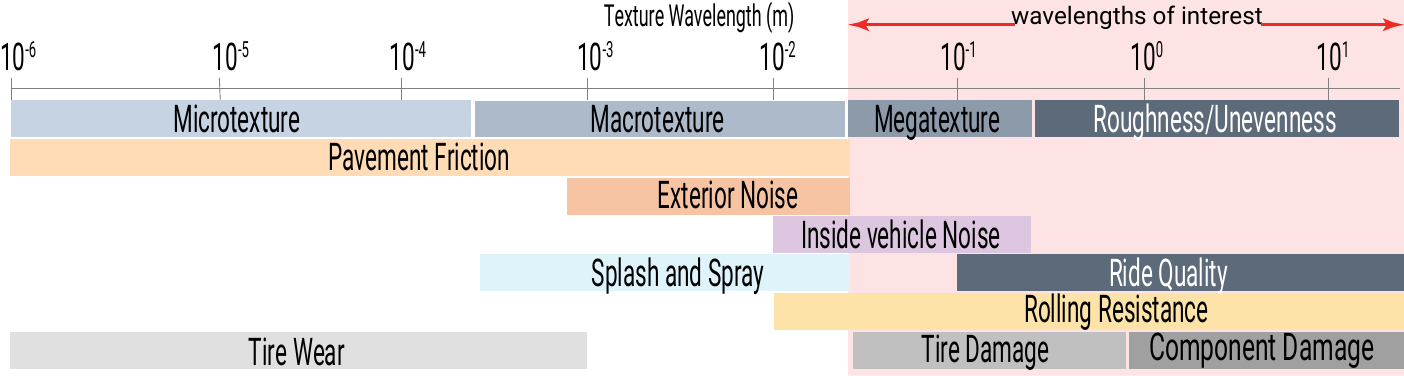}
 \vspace{-1em}
  \small \caption{Different wavelengths generated by various road damage types and their impact on a vehicle}
  \label{fig:wavelength}
  \end{center}  
\end{figure}

The roughness of the road, which is the main maintenance indicator of the road transport infrastructure, is characterized by the surface texture.
The road surface texture is divided into different categories based on the wavelength of the texture. 
Figure~\ref{fig:wavelength} shows different texture categories and how they affect the vehicle/tire interaction. 

\begin{figure}[ht]
	\centering
	  \includegraphics[keepaspectratio=true, width =0.46\textwidth]{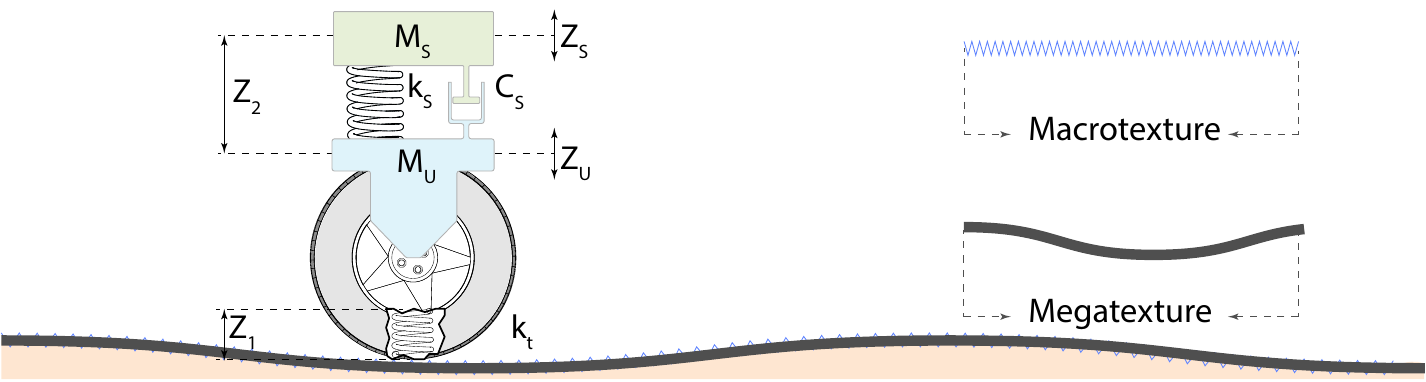}
				\small\caption{Quarter car model and effect of micro- and mega- textures on the car}
				\label{fig:Q_car}
				
	\end{figure}

While micro-texture is essential for the tire footprint to grip the asphalt, mega-textures are indication of pavement wear and damage. 
The vehicle vibrations are caused by mega-textures affecting the suspension and tire loads.
Mega-textures have wavelengths of 5cm or more. 
The ride quality is affected by the mega-textures with wavelength above 1m. Figure~\ref{fig:Q_car} shows the effects of micro- and mega- textures on a quarter car model.

The International Roughness Index (IRI) is a mathematical model that calculates the roughness based on the response of the sprung and unsprung mass of the quarter of the car. 

Our road roughness analysis service \cite{RoVI_seraj} calculates the roughness corresponding to the IRI by utilizing a set of adaptive signal decompositions, from which only the vibrations with wavelengths corresponding to the roughness are extracted. 
Because IRI defines the ratio between the overall vertical deviations and the profile length, the service allows creation of specific profile lengths.   

\subsubsection{Track geometry analysis}

\begin{figure}[ht]
	\centering
	  \includegraphics[keepaspectratio=true, width =0.46\textwidth]{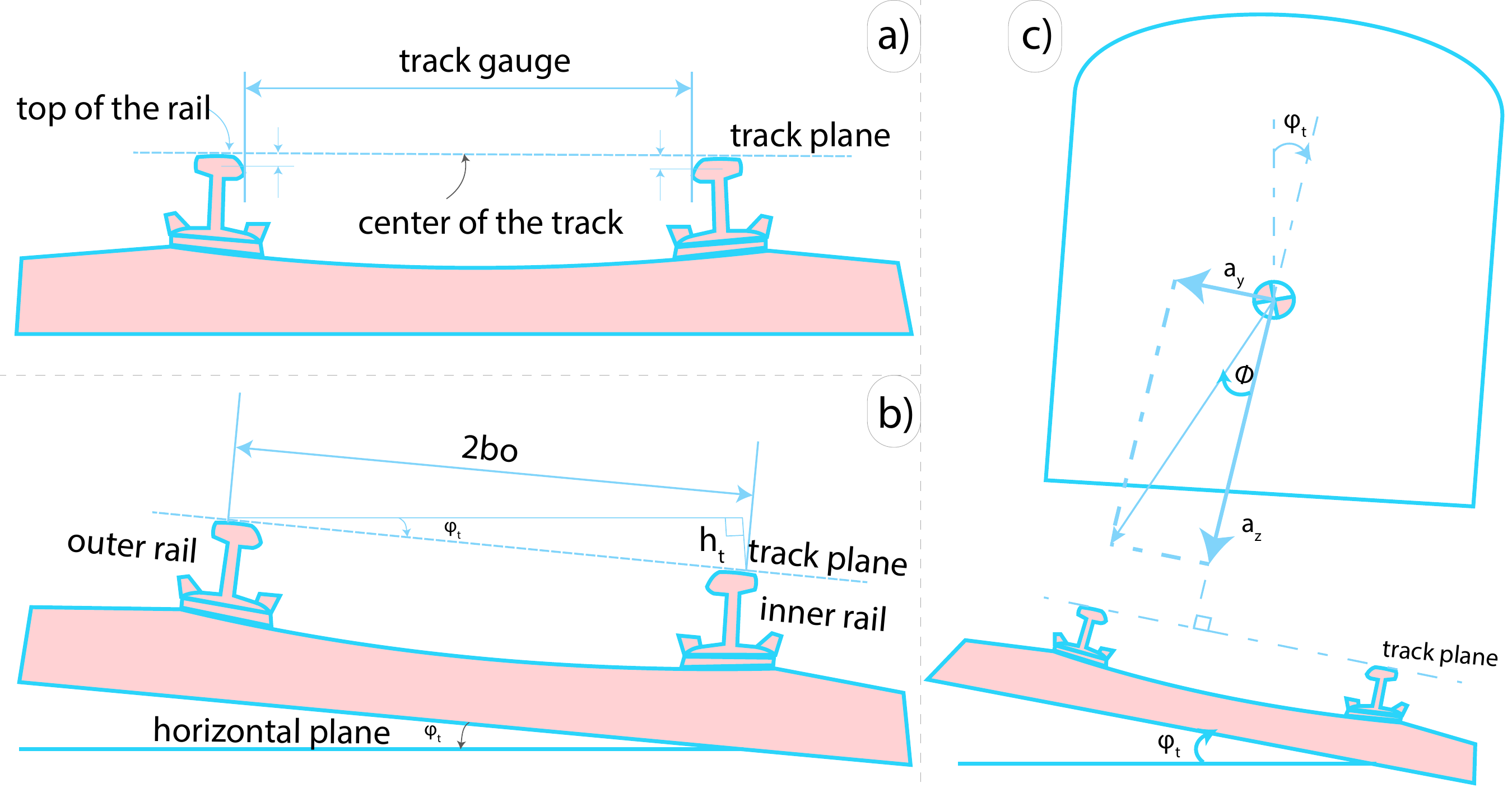}
				\small\caption{Train track gauge and cant}
				\label{fig:track}
				
	\end{figure}

Railroad tracks consist of two parallel metallic rails spaced from each other with a certain width called \textit{GAUGE}. As shown in Figure~\ref{fig:track}.a, the standard gauge width is 1435~mm. 
The locomotion of the trains is possible due to the friction between their metallic wheels with the rails. 
Whenever a train enters a curve, the track should provide a certain degree of slope for the train to resist the lateral forces. 
To provide the necessary slope, the outer rails should be placed higher than the inner rail. 
The difference between the two rails in the curvature is called \textit{CANT}. The angle of the cant is calculated using Equation~\ref{eq:artan1}, where $h_t$ is the rail elevation and $2b_0$ (as shown in Figure~\ref{fig:track}.b is the width between the center of the rails around 1500 mm. Cant is calculated when the coach enters and leaves the curve as well as all the twists for different wavelength \cite{RoVI_seraj}.

\begin{equation}
\label{eq:artan1}
   \varphi _{t}=asin\frac{h_{t}}{2b_{0}}
\end{equation}

In the case of passing through a curve, the train coach is exposed to two accelerations, i.e., (i) lateral acceleration $a_y$ parallel to the track plane and (ii) vertical acceleration $a_z$ perpendicular to the track plane.
 All the information regarding track geometry (cant, twist, curvature) as well as vertical profile can be measured and derived from the accelerometer and gyroscope readings, as well as the location of track segments from the GPS chip.
 The service calculates all the roll and pitch angles on the railroad track by differentiating the curvatures based on their length.
 This allows to classify curvatures based on their wavelength.

\section{System architecture}
\begin{figure}[ht]
	\centering
	  \includegraphics[keepaspectratio=true, width =0.46\textwidth]{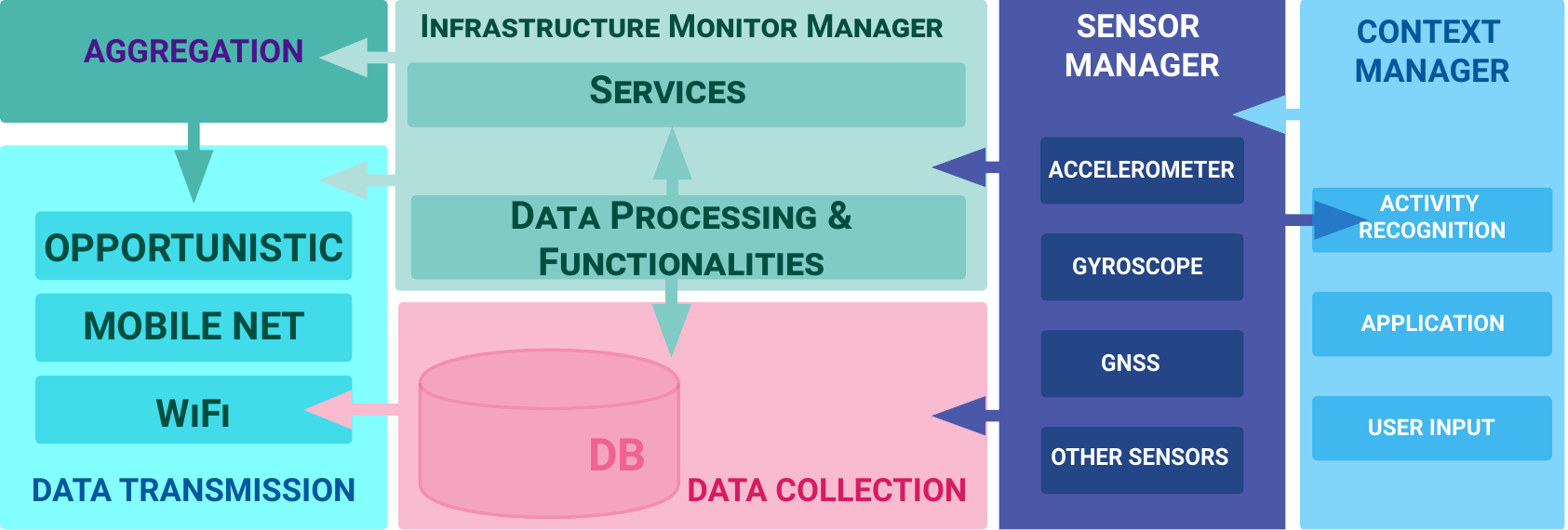}
				\small\caption{System architecture diagram}
				\label{fig:sysarch}
	\end{figure}

This section explains the architectural modules of our crowd sensing-based system for transport infrastructure monitoring using smartphones.
Most of the architectural modules share different services to decrease the architecture complexity and computation / communication overhead that can be overwhelming and minimize the benefits provided by the smartphone platform. 
Figure~\ref{fig:sysarch} shows the system architecture, showing that it consists of the following main modules:

\begin{itemize}
\item Context Manager module allows controlling various types of transport infrastructures and deciding how the sensor data will be processed.
\item Sensor Manager module, which manages and controls access to the sensing components. 
\item Infrastructure Monitor Manager module will process and calculate maintenance indicators based on the infrastructure type selected by the Context Manager component.
\item Aggregation module permits aggregation of the maintenance indicators and their location in situations when selected infrastructure segments are periodically monitored by the same device. This will further reduce the computation complexity on the cloud/server side.
\item Data Transmission module to  assist the system to communicate with the cloud as well with other smartphones opportunistically, utilizing WiFi and GPRS. 
\end{itemize}

In what follows we explain the above-mentioned modules.
\subsection{Context manager module}
Participants of the transport infrastructure monitoring can utilize a variety of transportation modes. Each transport infrastructure necessitates certain sensors, functionality and services to be used. Context manager module is responsible for the management of the transportation mode context. Monitoring features can be contextualized in three ways:
\begin{itemize}
\item User input, when the participant based on his needs selects the type of the infrastructure being monitored.
\item Application, when the proposed framework services are being used to develop or input information to other specific systems or platforms. In such cases the context is managed or predefined by exterior sources.
\item Activity recognition, when the device itself recognizes the transportation mode utilizing the input from Sensor manager and appoints the corresponding services according to the infrastructure being used. 
\end{itemize}

\subsection{Sensor manager module}
Access to the sensing components is provided by the smartphone OS. Although the OS allows arbitrary sampling rates, there is no guarantee that the sampling rate will be correct. Lack of a sensor will result in service limitations, for example the lack of gyroscope will provide limited results regarding vehicle orientation which will consequently lead to incorrect angle calculations (as the angles will be calculated by the accelerometer axis tilt).
However, the system provides the sensing requesting a list of available sensors. 
Once the sensor list is available and the sensors of interest are present, the system will investigate the maximum rate at which a sensor can acquire data.
Sensing allows the system to tap different sensors, even those are not directly related to the infrastructure state. For example light sensor can determine the light intensity providing environmental characteristics of the infrastructure.

While the data is being processed, the raw data can also be stored in a local database and sent to the cloud whenever a WiFi connection is available.As mentioned earlier, the raw sensor data is affected by uncertainties and noise associated with sensor hardware, environmental factors, and infrastructure maintenance indicators.
Building algorithms and implementing services based only on the sensors raw data only will provide incoherent results.  
An important factor that will rule the data management and handling is the sensor heterogeneity and varying sampling rates.
Data handling requires managing the system buffer. If a buffer is designed with a certain length, it will not work for higher or lower sampling rates, for example a 250 sample buffer corresponds to 2.5 sec with 100Hz but it can be less or more with lower or higher sampling rates.
This introduced an extra overhead for data handling to accommodate elastic buffers to satisfy both the application and smartphone environment requirements.

\subsection{Infrastructure monitor manager module}
Although the system performs a real-time calculation of the infrastructure maintenance indicators, the raw data contains other useful information and are collected for further processing in the cloud.
The data can be collected at maximum sensor sampling rate or as decided by the application.
The collected data are then sent to the cloud, using inexpensive data transfer options such as WiFi.

From the engineering point of view,  signals carry both useful and unwanted information. The distinction between useful and unwanted information requires the signal processing to be application dependent~\cite{ingle_digital_2012}.

Signal processing techniques allow to extract or enhance useful information from a stream of adverse information.
Altogether, signal processing can be described as an operation designed for extracting, enhancing, storing, and transmitting useful information. 
Inertial signals collected inside a running vehicle can be inspected by first analyzing the cause of vibration.
To this end, Infrastructure monitoring module is composed by two submodules:
\begin{inparaenum}
        \item Data processing \& functionalities
    \item Services
\end{inparaenum}

\subsection{Data processing module}
Considering that measuring each specific infrastructure segment will result in huge amount of sensor data with considerable redundant information, it is more appropriate to compute derived values representing valuable information. To this end, feature extraction is a method to reduce dimensionality of a dataset.
One should note that sensor data is streaming and arrives every $dt$. However the statistical methods require availability of the entire data to be able to calculate valuable information. These reasons motivate us to use feature extraction techniques.

\begin{figure}
\centering
  	  \includegraphics[keepaspectratio=true, width =0.46\textwidth]{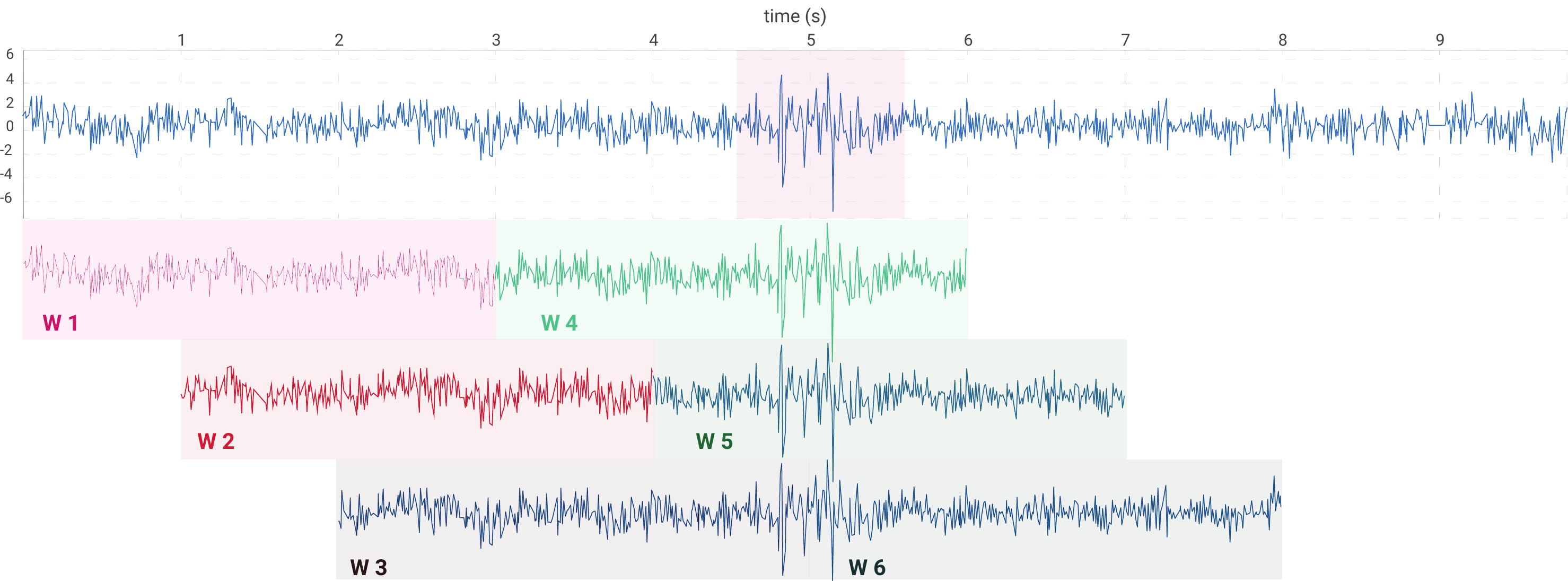}
				\small\caption{ Framing a 10 second signal into 3 second windows with a $\frac{1}{3}$d overlap}
				\label{fig:windowing}
\end{figure}

For the purpose of feature extraction, we frame the signal into windows containing only finite length part of the signal. In order to avoid situations in which the signal is framed exactly at the point of interest (occurrence of a relevant monitoring event) and consequently loosing the full information about that precise event, we allow the frames to overlap with each other. Figure~\ref{fig:windowing} shows a 10 second signal representing a pothole event at the $5_{th}$ second.
After framing the signal, we calculate the features related to the pothole event based on data form the window $W4$ and $W5$. If the frames do not overlap (for instance framing the signal into $W3$ and $W6$), one can see that the information about the pothole is only partially available, leading to missing part of information about properties of the event.Each window can be described as a feature vector and the entire signal as a feature matrix.

Consider a signal $S$ framed with $n$ window $w$ with a length of $m$ data points $d$.
From each window a set of $k$ features $f$ are extracted, resulting in a feature vector $V$ for each window of the form:
\begin{equation}
V_n=\begin{pmatrix}
 f_1^n, f_2^n,  ..., f_k^n 
\end{pmatrix}
\end{equation}

The whole signal $S$ than can be represented as feature matrix $M$ of the form:
\begin{equation}
M=\begin{bmatrix}
f_1^1 & f_2^1 & ... & f_k^1 \\ 
f_1^2 & f_2^2 & ... & f_k^2 \\ 
... & ... & ... & ... \\ 
f_1^n & f_2^n & ... & f_k^n 
\end{bmatrix}
\end{equation}

\subsubsection{Feature library}
\begin{figure}
\centering
  	  \includegraphics[keepaspectratio=true, width =0.46\textwidth]{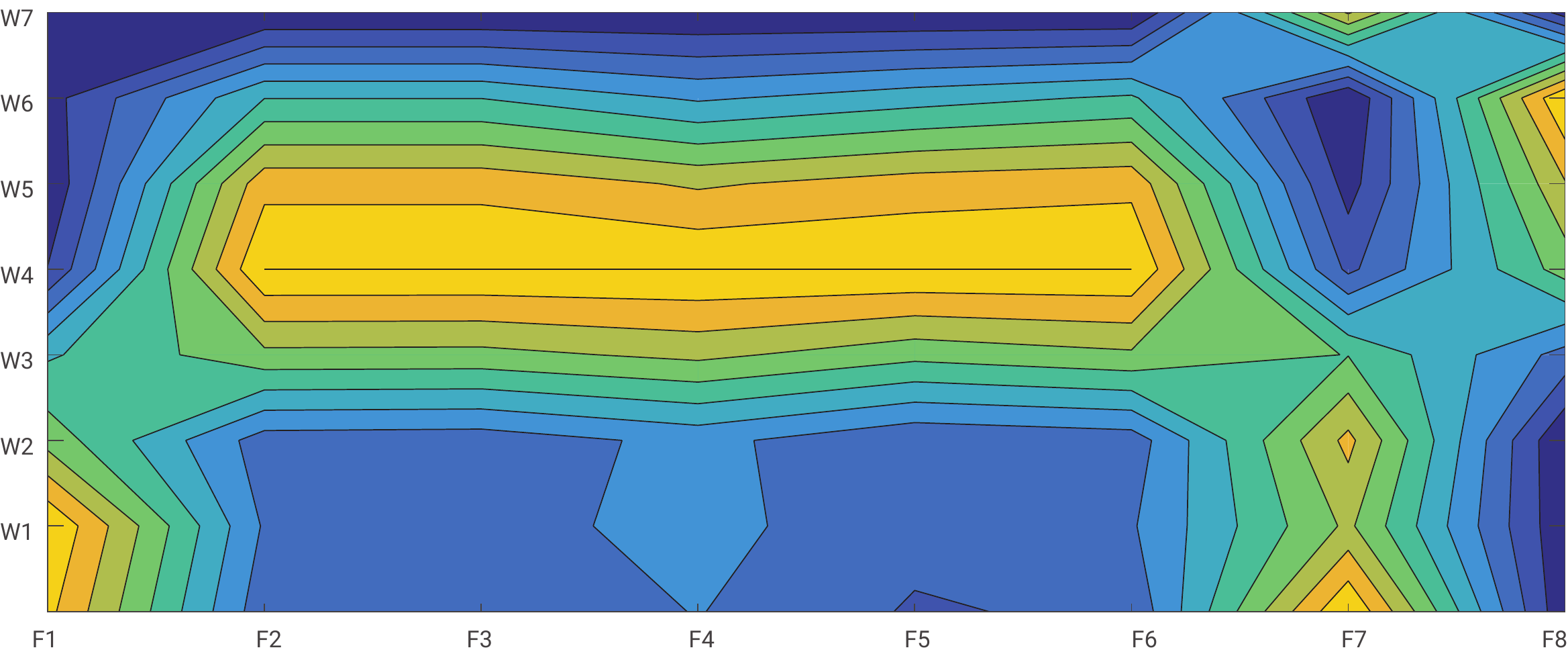}
				\small\caption{ A contour plot of normalized features for 7 windows of the signal in Figure~\ref{fig:windowing} }
				\label{fig:features}
\end{figure}

We provide a library of reference features to be used for extracting features based on the context scenario of the infrastructure to be monitored.
The library consists of the following features:
\begin{inparaenum}[i)]

    \item Mean, describes statistically the expected values of the measured central tendency.
    
    \item Median Absolute Deviation (MAD) is a measure of scale based on the median of the absolute deviations from the median of the distribution. The median deviation compared to standard deviation is  less efficient  as a measure of scale for normal distributions, but more robust for distributions with heavier tails~\cite{everitt_encyclopedia_2005}.
   
    \item RootMeanSquare (RMS) is the mean of the squared values of the sample set.
    
    \item Variance (VAR), which is defined as the mean of the squared deviations of a set of numbers about their mean. As with other summary statistics that rely equally on all the numbers in the set, the variance can be severely affected by extreme scores~\cite{everitt_encyclopedia_2005}. 
    \item Standard deviation (SD) measures the spread or dispersion, is defined as the (positive) square root of the variance. The SD is a way of putting the mean of a dataset in context and facilitates comparison of the distributions of several samples by showing their relative spread~ \cite{everitt_encyclopedia_2005}.
    
    \item Energy (L2-norm) calculated based on the total size or length of all vectors in a vector space.
    
    \item Skewness, which is a measure of symmetry, or more precisely, the lack of symmetry. A distribution, or data set, is symmetric if it looks the same to the left and right of the center point. The skewness for a normal distribution is zero, negative values for the skewness indicate data that are skewed left and positive values for the skewness indicate data that are skewed right~\cite{_nist/sematech_????}.
    
    \item Kurtosis, describes whether the data are heavy-tailed or light-tailed relative to a normal distribution, data with high kurtosis tend to have heavy tails, or outliers~\cite{_nist/sematech_????}. 
    
    \item Peak2Peak calculates the difference between the maximum and minimum values of the signal.
    
    \item Peak2RMS calculates the ratio of the siganls largest absolute value to the RMS value of signal.
\end{inparaenum}
For example when the above normalized features are calculated for the 7 windows of the signal in Figure~\ref{fig:windowing}, can be observed that the anomalous windows $W4$ and $W5$ are prominent through the feature set as shown in Figure~\ref{fig:features}.

\subsubsection{Signal transformation}
Since every vibration is caused by a specific wavelength, caused by a spinning component on the vehicle or a vertical displacement on the infrastructure, it makes sense to use a transform-based signal processing method~\cite{vaseghi_advanced_2006}.

The idea behind transform-based signal processing is to represent the signal as a composite of underlying signals such as sinusoid or wavelet. This form of decomposition of complex signals into basic ones allows a simpler interpretation and analysis.
This category of transform-based signal processing include the following transform methods that decomposes a function of time into composing frequencies: 
\begin{inparaenum}
	\item Fourier transform 
	\item Wavelet transformation
	\item Hilbert-Huang transform
\end{inparaenum}
\subsubsection{Fourier transform}
 Fourier transform is a widely used analytic method for signal frequency analysis in particular for vibration analysis. For the Fourier transform, the base functions are the complex oscillations $b_\omega:=exp(i\omega t)$, where t is the time axis and $\omega$ is the single frequency parameter that determines the basis function in the family.  
\begin{equation}
    F{x(t)}(\omega)=\left \langle b(\omega,t),s(t) \right \rangle=\int_{-\infty}^{\infty}exp(-i\omega \tau)s(\tau)d\tau
\end{equation}

The drawback of the Fourier transform is the loss of the time information of the transformed non-stationary transient signal.
Short Time Fourier Transform (STFT) overcomes this drawback by windowing the signal into short-enough segments to be considered stationary.     However, not all the signals are stationary. A signal is considered non-stationary when its frequency or spectral content change with respect to time. The base function becomes $b_\omega:=w(t-t_0) exp(i\omega t)$, where $w(t)$ is the window function that vanishes outside some intervals and $(\omega,t_0)$ is the time-frequency coordinates of the base function in the family, making the inner product in the form of:

\begin{equation}
    STFT{x(t)}(\omega,t_0)=\int_{-\infty}^{\infty}w(\tau)*exp(-i\omega \tau)s(\tau)d\tau
\end{equation}

Thus, STFT can outline the signal into a function of frequency and time. The result of this transform can also be regarded as a filter-bank with bandpass filters that have the Fourier transform as the window $w(t)$ as frequency response, but shifted to the center frequency $\omega$. All filters therefore have the same bandwidth.

Although the STFT resolves the time frequency resolution, still actual trade-off between time and frequency is determined by the choice of the window function. 
With regard to the profile lengths, the frequency of the wavelength is proportional to the velocity, meaning that the resolution will not be constant throughout the analysis. Thus, this makes the window size decision a challenging task.    
\subsubsection{Wavelet transform} 
Wavelet transform is a  multi-resolution transform where a signal is decomposed and described as a combination of elementary waves called wavelets of different duration. The basis function of the wavelet transform is basically the coefficients of the contractions and dilutions of the wavelet. Thus, this allows non-stationary events of various duration in a signal to be identified and analyzed. This is quite important considering that non-stationary events of the same category in a infrastructure signal varies in time corresponding to the speed of travel.

Recovering the signal only from the waveform related to the infrastructure geometry requires the implementation of signal decomposition methods capable of providing a time frequency representation of the signal. 
The frequency representation is obtained through transformation of the time series signal. The output of this transform is the inner product of a family of basis functions with the signal.

The wavelet transform resolves the constant bandwidth constraint by adapting the window size to the frequency. This happens in a specific scale invariant way that does not even need the complex modulation anymore. 
The generic base function becomes a wavelet that is localized and oscillates and it also has zero mean (i.e. the integral over the complete space is 0)~\cite{mallat_wavelet_2009}. 
When the wavelet scales in time, the oscillation frequency changes as well. This leads to controlling the localization and oscillation with a single linking parameter. 
The family of base functions becomes, 
\begin{equation}
b(\sigma,t_0)(t)=w(\frac{t-t_0}{\sigma})
\end{equation}
where $w$ is the original wavelet and $\sigma$ is the scale parameter. The inner product becomes:
\begin{equation}
    W{s(t)}(\sigma,t_0)=\int_{-\infty}^{\infty}w(\frac{t-t_0}{\sigma})*s(t) dt
\end{equation}

The transformation provides a two dimension parametric result that can be seen as a filter bank. However, the filter bandwidth is proportional to the center frequency of the bands. Thus, the time-frequency plane is partitioned non-uniformly.

The Discrete Wavelet Transform (DWT) analyzes the signal at different frequency bands with different resolutions by decomposing the signal into a coarse approximation and detail information. DWT employs two sets of functions, i.e., (i) scaling functions and (ii) wavelet function, which are associated with low pass and high pass filters, respectively. 
The signal decomposition into different frequency bands is obtained by successive high and low pass filtering the original signal. 

\begin{figure}[ht]
	\centering
	  \includegraphics[keepaspectratio=true, width =0.46\textwidth]{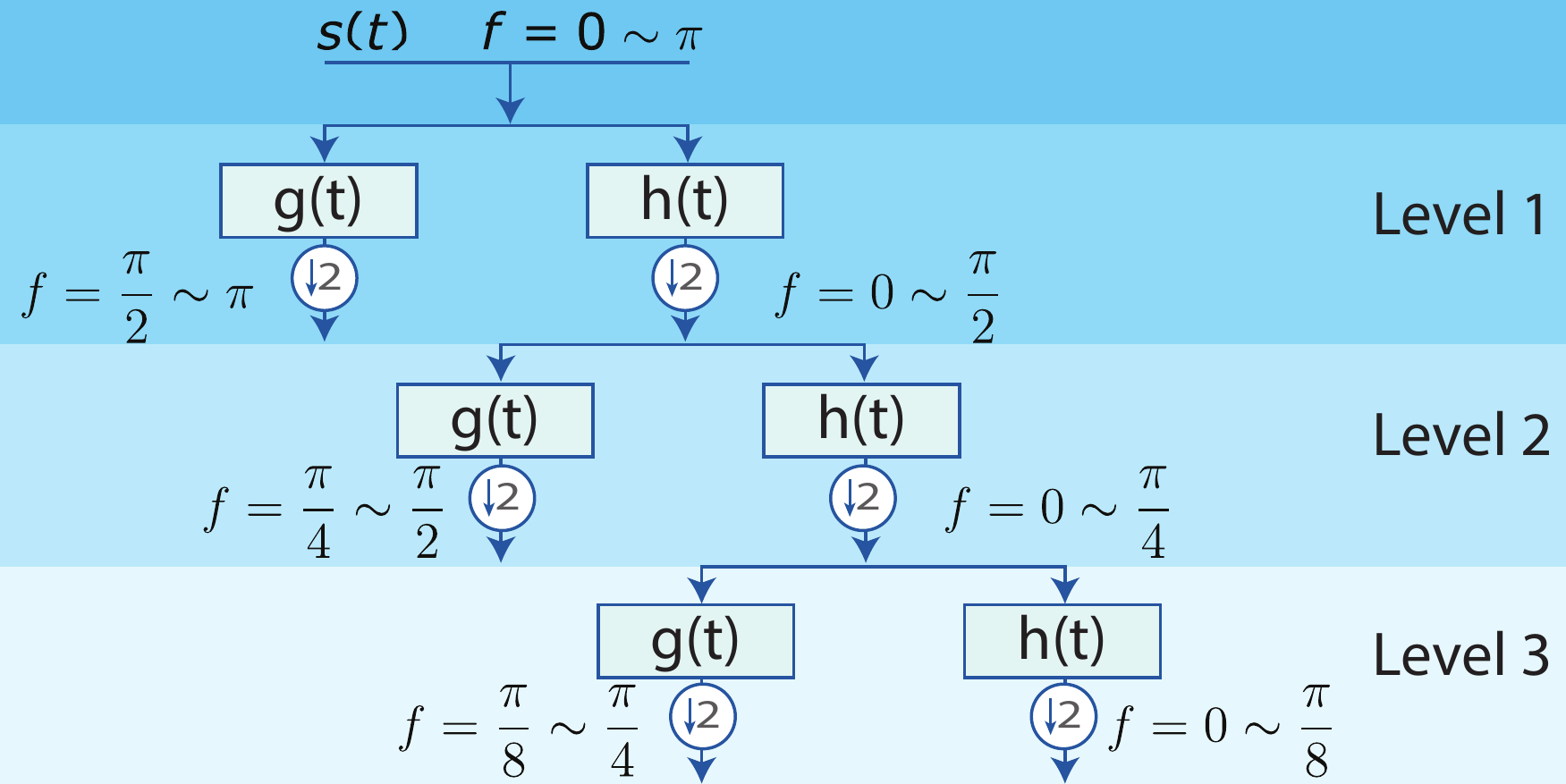}
				\small\caption{ Discrete Wavelet Transform and subband coding algorithm}
				\label{fig:dwt}
	\end{figure}

The original signal $s(t)$ is first passed through a halfband high pass filter $g(t)$  and a low pass filter $h(t)$. This is shown in Figure~\ref{fig:dwt}. After filtering, half of the samples can be eliminated according to the Nyquist rule, since the signal now has the highest frequency of $\frac{\pi}{2}$ radians instead of $\pi$. The signal can be sub-sampled by 2, simply by omitting every other sample.
\subsubsection{Hilbert-Huang transform} 
Hilber- Huang transform is another method for time-frequeny analysis, the empirical mode decomposition–Hilbert-Huang transform (EMD-HHT), is applicable to both nonstationary and nonlinear signals. EMD-HHT method requires two steps in analyzing the data. The first step is to decompose time series data into a number of intrinsic mode functions (IMF), The second step is to apply the Hilbert transform to the decomposed IMFs and construct the energy–frequency–time distribution, designated as the Hilbert spectrum. The presentation of the final results of the time–frequency analysis is
similar to the wavelet transform method, which is a spectrogram.
In HHT the decomposition of a time series $H(t)$ into  IMF is achieved by extracting the upper envelope of the local maximas and the lower envelope of the local minima. The mean of the envelopes is designated as $m_1(t)$. Extracting the mean from the $H(t)$ results in the first component $h_1(t)$. This procedure can be repeated $k$ times, until $h_k(t)$ is an IMF, thus $h_k(t) =c_1(t)$. $c_1(t)$ contains the shortest period component of the signal as all the low frequency components represented by the extracted envelope are removed. 
The $c_1(t)$ is removed from $H(t)$ and the process is repeated with the residue $r_1(t)$. This process continues as long as either the $c_n(t)$ or $r_n(t)$ become smaller than a predetermined value or the residue $r_n(t)$ becomes a monotonic function with no extractabel IMF \cite{huang_hilbert}.
\begin{equation}
    H(t)=\sum_{i=1}^{n}C_1(t)+r_n(t)
\end{equation}
Considering the properties of the signals measured on the infrastructure being linear non-stationary signals and the aforementioned properties of these signal transformation methods, we chose the Stationary Wavelet Transformation (SWT) for the signal processing service of the system.
The service allows to analyse the signal by choosing the decomposition levels to reconstruct the signal only on certain frequency bands and extract the statistical features for each signal partition.

\subsection{Aggregation module}
Compared to the dedicated sensors of high-end infrastructure monitoring measurement vehicles, the measurement accuracy of the smartphones are much lower. 
However, this inaccuracy can be compensated by the large number of repetitive measurements made on a daily basis by large number of active participants. This guarantees the required spatial and temporal resolution for determining the infrastructure state.

The aggregation module addresses the following challenges posed by the crowd-sourced nature of measurements: 
\begin{inparaenum}

    \item Sparse and uncoordinated measurements: smartphone owners measure infrastructure segments randomly without having a priori knowledge about the track being measured.

    \item Heterogeneity of devices: different types of smartphones with different sampling rates, processing power, operating systems, and sensor accuracy will be used.

    \item Unequal spatial and temporal resolutions: temporal and spatial sampling frequency will be different and vehicular speed and GPS inaccuracy will increase the latency. 
\end{inparaenum}

To achieve the required accuracy in both space and time, computational geometry method such as the Voronai Diagrams or the Delaunay Triangulation ($DT$) are used locally and in cloud back-end, described in \cite{seraj_aggregation} and implemented in \cite{jansen_e-sight_2020}.
Pairing measured segments together and updating the aggregate position with new contributions permits an accurate and efficient map-matching.

Map-matching challenge consists in combining the streaming data from the smartphones in space and time and translating the uncertain GPS measurements into an exact geographical location of transport infrastructure segment. 

The module also addresses the problem of spatio-temporal data measurements, proposing a time-machine approach, in which the evolution of the infrastructure segment state is updated using a weighted function that dilute previous measurements and weight up the new ones, yielding an accurate time resolution state.

\subsection{Data transmission module}
This module assists the system to communicate with other smartphones opportunistically, utilizing WiFi and GPRS. Although the system performs a real-time calculation of the infrastructure maintenance indicators, the raw data contains other useful information that can be extracted using data mining techniques. Collecting this data for further processing is therefore a justified action.
The data can be collected at maximum sensor sampling rate or as decided by the application.
The collected data are then sent to the cloud, using inexpensive data transfer options such as WiFi.
\subsubsection{Data connectivity requirements}
Data connectivity is a fundamental requirement for crowd sensing-based systems, as the paradigm is based on in-situ contribution of individuals.
Due to the fact that for each segment of the transport infrastructure, only the geo-location (coordinates) and the computed maintenance indicators need to be transmitted, a wide bandwidth communication channel is not needed.

However, from a big data perspective, transmission and availability of other intermediary computations or the raw sensed data may be beneficial for further analysis. 
Therefore, availability of different types of data transmission may facilitate and reduce transmission bottlenecks. 
Modern smartphones provide mobile network data connectivity (via GPRS) as well as WiFi and Bluetooth, which can be exploited for use in transport infrastructure monitoring. 
\subsubsection{Cooperative opportunistic data dissemination}
An opportunistic vehicle to vehicle network, as described in \cite{Turkes1022}, facilitates an increase in accuracy for the anomaly detection service by collaborating and exchanging information regarding transport infrastructure indicators and features without requiring the smartphones to communicate with a central cloud service. Its execution is prioritized based on severity of anomaly being detected by smartphones.
The network relies on  the communication range of smartphone's WiFi and the WiFi chip switching between the hot-spot mode to disseminate smartphone data and transport infrastructure segment information and the client mode to receive the information utilizing the SSID packet to encode the road information.
Because the SSID length is limited to 32 bytes only, the data packets need to be encoded in such a way that it encloses all important information such the location, type, number, and nature of the infrastructure anomalies a given road length of are radius.

\section{Summary}
In this paper, we have presented requirements and architecture of our crowd sensing-based infrastructure monitoring system. The system relies on the sensing, communication, and processing capabilities of the modern smartphones as well as sophisticated data processing, classification and clustering algorithms. 
The processing scheme of the system can be implemented on top of the smart IoT device allowing a wide range of deployment. 
The system architecture encapsulated required modules and components to support and monitor different infrastructure types, their associated geometry indicators, users, authorities and responsible engineers. Short description of principles and concepts of these modules presented in this chapter will be followed with details description of our designed methods and algorithms to instantiate the components in the upcoming chapters.


\bibliographystyle{IEEEtran}
\bibliography{main.bib}

\begin{thebibliography}{10}
\providecommand{\url}[1]{#1}
\csname url@samestyle\endcsname
\providecommand{\newblock}{\relax}
\providecommand{\bibinfo}[2]{#2}
\providecommand{\BIBentrySTDinterwordspacing}{\spaceskip=0pt\relax}
\providecommand{\BIBentryALTinterwordstretchfactor}{4}
\providecommand{\BIBentryALTinterwordspacing}{\spaceskip=\fontdimen2\font plus
\BIBentryALTinterwordstretchfactor\fontdimen3\font minus
  \fontdimen4\font\relax}
\providecommand{\BIBforeignlanguage}[2]{{%
\expandafter\ifx\csname l@#1\endcsname\relax
\typeout{** WARNING: IEEEtran.bst: No hyphenation pattern has been}%
\typeout{** loaded for the language `#1'. Using the pattern for}%
\typeout{** the default language instead.}%
\else
\language=\csname l@#1\endcsname
\fi
#2}}
\providecommand{\BIBdecl}{\relax}
\BIBdecl

\bibitem{Miller:2009te}
T.~Miller and E.~Zaloshnja, ``On a crash course: The dangers and health costs
  of deficient roadways,'' Transportation Construction Coalition, Tech. Rep.,
  2009.

\bibitem{kinney_development_1986}
J.~L.~M. Kinney, ``Development of a pavement management system for the city of
  indianapolis,'' in \emph{Road School, Purdue University Proceedings of the
  annual road school}, ser. Road school, Purdue, 1986, pp. 50--63.

\bibitem{schut_responsible_2000}
P.~Schut, T.~de~Bree, and G.~Fuchs, ``Responsible pavement management,'' in
  \emph{First European Pavement Management System: Conference-Procidings and
  final program}, 2000.

\bibitem{ProakisManolakis200604}
\BIBentryALTinterwordspacing
J.~G. Proakis and D.~K. Manolakis, \emph{Digital Signal Processing (4th
  Edition)}, 4th~ed.\hskip 1em plus 0.5em minus 0.4em\relax Pearson, 4 2006.
  [Online]. Available: \url{http://amazon.com/o/ASIN/0131873741/}
\BIBentrySTDinterwordspacing

\bibitem{AGPS:_2002}
J.~{LaMance}, J.~{DeSalas}, and J.~Jarvinen, ``Innovation: Assisted {GPS:} a
  low-infrastructure approach,'' \emph{{GPSWorld}}, no. March, pp. 46--51,
  2002.

\bibitem{chandola_anomaly_2009}
\BIBentryALTinterwordspacing
V.~Chandola, A.~Banerjee, and V.~Kumar, ``Anomaly detection: A survey,''
  \emph{ACM Comput. Surv.}, vol.~41, no.~3, pp. 15:1--15:58, Jul. 2009.
  [Online]. Available: \url{http://doi.acm.org/10.1145/1541880.1541882}
\BIBentrySTDinterwordspacing

\bibitem{Seraj:angle}
\BIBentryALTinterwordspacing
F.~Seraj, K.~Zhang, O.~Turkes, N.~Meratnia, and P.~J.~M. Havinga, ``A
  smartphone based method to enhance road pavement anomaly detection by
  analyzing the driver behavior,'' in \emph{Adjunct Proceedings of the 2015 ACM
  UbiComp/ISWC' 2015 International Symposium on Wearable Computers}, ser.
  Adjunct.\hskip 1em plus 0.5em minus 0.4em\relax ACM, 2015, pp. 1169--1177.
  [Online]. Available: \url{http://doi.acm.org/10.1145/2800835.2800981}
\BIBentrySTDinterwordspacing

\bibitem{RoVI_seraj}
F.~{Seraj}, N.~{Meratnia}, and P.~J.~M. {Havinga}, ``Rovi: Continuous transport
  infrastructure monitoring framework for preventive maintenance,'' in
  \emph{2017 IEEE International Conference on Pervasive Computing and
  Communications (PerCom)}, 2017, pp. 217--226.

\bibitem{ingle_digital_2012}
V.~K. Ingle and J.~G. Proakis, \emph{Digital signal processing using
  {MATLAB}}.\hskip 1em plus 0.5em minus 0.4em\relax Cengage Learning, 2012.

\bibitem{everitt_encyclopedia_2005}
B.~Everitt and D.~C. Howell, Eds., \emph{Encyclopedia of statistics in
  behavioral science}.\hskip 1em plus 0.5em minus 0.4em\relax John Wiley \&
  Sons, 2005.

\bibitem{_nist/sematech_????}
``{NIST}/{SEMATECH} e-handbook of statistical methods,''
  \url{http://www.itl.nist.gov/div898/handbook/}, 2012.

\bibitem{vaseghi_advanced_2006}
S.~V. Vaseghi, \emph{Advanced digital signal processing and noise reduction},
  3rd~ed.\hskip 1em plus 0.5em minus 0.4em\relax Wiley, 2006.

\bibitem{mallat_wavelet_2009}
S.~G. Mallat, \emph{A wavelet tour of signal processing: the sparse way},
  3rd~ed.\hskip 1em plus 0.5em minus 0.4em\relax Amsterdam ; Boston:
  Elsevier/Academic Press, 2009.

\bibitem{huang_hilbert}
N.~Huang and N.~Attoh-Okine, \emph{The Hilbert-Huang Transform in
  Engineering}.\hskip 1em plus 0.5em minus 0.4em\relax Taylor \& Francis, 2005.

\bibitem{seraj_aggregation}
F.~Seraj, N.~Meratnia, and P.~Havinga, \emph{\BIBforeignlanguage{English}{An
  aggregation and visualization technique for crowd-sourced continuous
  monitoring of transport infrastructures}}.\hskip 1em plus 0.5em minus
  0.4em\relax United States: IEEE, 5 2017, pp. 219--224.

\bibitem{jansen_e-sight_2020}
M.~Jansen and F.~Seraj, ``\BIBforeignlanguage{en}{e-{Sight}: {Real}-time cloud
  platform for visualizing edge transport infrastructure information},'' in
  \emph{\BIBforeignlanguage{en}{2020 {IEEE} {International} {Conference} on
  {Pervasive} {Computing} and {Communications} {Workshops} ({PerCom}
  {Workshops})}}.\hskip 1em plus 0.5em minus 0.4em\relax IEEE, Mar. 2020, pp.
  1--6.

\bibitem{Turkes1022}
O.~Turkes, F.~Seraj, H.~Scholten, N.~Meratnia, and H.~Paul J.~M., ``An ad-hoc
  opportunistic dissemination protocol for smartphone-based participatory
  traffic monitoring,'' in \emph{Vehicular Technology Conference (VTC Fall),
  2015 IEEE 82nd}, September 2015.

\end{thebibliography}

\end{document}